\documentclass[prl,twocolumn,showpacs]{revtex4}
\usepackage{amsmath}
\usepackage{amssymb}
\usepackage{graphicx}
\usepackage{bm}
\usepackage{dcolumn}

\newcommand{\bpm}{\begin{pmatrix}}
\newcommand{\epm}{\end{pmatrix}}

\begin{document}

\title{Longitudinal interlayer magnetoresistance in quasi-2D metals}
\author{P. D. Grigoriev}
\email{grigorev@itp.ac.ru}
\affiliation{L. D. Landau Institute for Theoretical Physics, Chernogolovka, Russia}
\date{\today }

\begin{abstract}
The longitudinal interlayer magnetoresistance $R_{zz}\left( B_{z}\right) $
is calculated in strongly anisotropic layered metals, when the interlayer
band width $4t_{z}$ is less than the Landau level separation $\hbar \omega
_{c}$. The impurity scattering has much stronger effect in this regime than
in 3D metals and leads to a linear longitudinal interlayer magnetoresistance
$R_{zz}\propto B_{z}$ in the interval $\hbar \omega _{c}>4t_{z}>>\sqrt{%
\Gamma _{0}\hbar \omega _{c}}$ changing to a square-root dependence $%
R_{zz}\propto B_{z}^{1/2}$ at higher field or smaller $t_{z}$. The crossover
field allows to estimate the interlayer transfer integral as $t_{z}\sim
\sqrt{\Gamma _{0}\hbar \omega _{c}}$. Longitudinal interlayer
magnetoresistance, being robust to the increase of temperature or long-range
disorder, is easy for measurements and provides a useful tool to investigate
the electronic structure of quasi-two-dimensional compounds.
\end{abstract}

\maketitle

The investigation of angular and field dependence of
magnetoresistance (MR) provides a powerful tool of studying the
electronic properties of various metals, including strongly
anisotropic layered compounds, such as organic metals (see, e.g.,
Refs. \cite{MQORev,OMRev,MarkReview2004,KartPeschReview} for
reviews), cuprate and iron-based high-temperature
superconductors,(see,
e.g., \cite%
{HusseyNature2003,AbdelNature2006,ProustNature2007,AbdelPRL2007AMRO,McKenzie2007,DVignolle2008,HelmNd2009,HelmNd2010,BaFeAs2011,Graf2012}%
) heterostructures\cite{Kuraguchi2003} etc. In layered quasi-two-dimensional
(Q2D) metals with at least monoclinic crystal symmetry the electron
dispersion in the tight-binding approximation is given by
\begin{equation}
\epsilon _{3D}\left( \mathbf{k}\right) \approx \epsilon _{2D}-2t_{z}\cos
(k_{z}d),  \label{ES3D}
\end{equation}%
where the 2D electron dispersion in magnetic field perpendicular to
conducting layers is quantized in Landau levels:
\begin{equation}
\epsilon _{2D}=\epsilon _{2D}\left( n\right) =\hbar \omega _{c}\left(
n+\gamma \right) ,  \label{E2D}
\end{equation}%
where the Landau level (LL) separation $\hbar \omega _{c}=\hbar eB/m^{\ast
}c $, $m^{\ast }$ is the effective electron mass, $n$ is the LL number, $%
\gamma \approx 1/2$, $k_{z}$ is out-of-plane electron momentum, and $d$ is
the interlayer spacing. If the interlayer transfer integral $t_{z}\gg \hbar
\omega _{c}$, the standard 3D theory of galvanomagnetic properties\cite%
{Abrik,Shoenberg,Ziman} can be applied, which is valid in the lowest order
of the parameter $\hbar \omega _{c}/t_{z}$. This theory predicts some
special features of MR in Q2D metals: the angular magnetoresistance
oscillations (AMRO)\cite{KartsAMRO1988,Yam,Yagi1990} and the beats of the
amplitude of magnetic quantum oscillations (MQO).\cite{Shoenberg}

In more anisotropic Q2D metals, when $t_{z}\gtrsim \hbar \omega _{c}$,
several new features appear, such as slow MR oscillations\cite{SO,Shub} and
the phase shift of MQO beats between transport and thermodynamic quantities.%
\cite{PhSh,Shub} These two effects are not described by the standard 3D
theory \cite{Abrik,Shoenberg,Ziman} because they appear in the higher orders
in the parameter $\hbar \omega _{c}/t_{z}$. The monotonic part of MR also
changes when $\hbar \omega _{c}/t_{z}\sim 1$. According to the standard
theory,\cite{Abrik} external magnetic field along the electric current leads
only to MQO but does not influence the monotonic (background) part of this
current.\cite{CommentMaltsev} However, the monotonic growth of interlayer MR
$R_{zz}$ with the increase of longitudinal magnetic field $B_{z}$ was
observed in various compounds as a general feature of Q2D metals.\cite%
{SO,Coldea,PesotskiiJETP95,Zuo1999,W3,W4,Incoh2009,Wosnitza2002,Kang} Its
theoretical description is the aim of the present paper.

In very anisoropic and dirty compounds with $t_{z}\ll \Gamma _{0},\hbar
\omega _{c}$, where $\Gamma _{0}=\hbar /2\tau _{0}$ and $\tau _{0}$ is
electron mean free time in the absence of magnetic field, this monotonic
growth of $R_{zz}\left( B_{z}\right) $ was attributed to the new "strongly
incoherent" mechanisms of interlayer electron transport, such as
metal-insulator transition and variable-range electron hopping with
exponential temperature and field dependence of $R_{zz}$,\cite{Gvozd2007} or
the interlayer hopping via local crystal defects with in-series metallic
intralayer transport.\cite{Incoh2009} The in-plane component of magnetic
field in the "strongly incoherent" regime leads to weaker MR compared to
coherent 3D theory.\cite{Incoh2009,Lundin2003,Ho} However, in the most
experiments the longitudinal interlayer magnetoresistance was observed
together with the pronounced AMRO and metallic-type temperature dependence,
which is inconsistent with "strongly incoherent" mechanisms of interlayer
electron transport, which do not conserve the in-plane electron momentum
during interlayer electron hopping. Recently it was shown,\cite%
{WIPRB2011,WIJETP2011,WIFNT2011} that at very weak interlayer coupling, $%
\hbar \omega _{c}\gg \Gamma _{0}\gg t_{z}$, the longitudinal interlayer
magnetoresistance has a square-root monotonic growth $R_{zz}\propto \sqrt{%
B_{z}}$ even within the coherent-tunnelling model. The angular dependence of
MR also changes in this limit,\cite{WIPRB2011} which contradicts the
previous common opinion\cite{MosesMcKenzie1999} that in the "weakly
incoherent" regime, i.e. at $\Gamma _{0}>t_{z}$, the interlayer
magnetoresistance does not differ from the coherent almost 3D limit $%
t_{z}\gg \Gamma _{0}$. In the present paper we calculate the longitudinal
interlayer magnetoresistance at $\Gamma _{0}\lesssim 4t_{z}<\hbar \omega
_{c} $, which generalizes the result of Refs. \cite{WIPRB2011,WIJETP2011}.
This extends the applicability of the square-root dependence $R_{zz}\propto
\sqrt{B_{z}}$ to much wider region $\hbar \omega _{c}>\sqrt{\Gamma _{0}\hbar
\omega _{c}}\gtrsim t_{z}$ and gives linear $R_{zz}\left( B_{z}\right) $
dependence in the interval $\hbar \omega _{c}>4t_{z}\gg \sqrt{\Gamma
_{0}\hbar \omega _{c}}$.

\medskip

At $2t_{z}>\Gamma _{0}$ one can use the 3D electron dispersion in Eq. (\ref%
{ES3D}). Then the interlayer electron conductivity can be evaluated at
finite temperature using the 3D Kubo formula,\cite{Mahan} which gives%
\begin{equation}
\sigma _{zz}=\int d\varepsilon \,\left[ -n_{F}^{\prime }(\varepsilon )\right]
\,\sigma _{zz}(\varepsilon ),  \label{sigmazz}
\end{equation}%
where the derivative of the Fermi distribution function\cite%
{CommentChemPotOsc}
\begin{equation}
n_{F}^{\prime }(\varepsilon )=-1/\{4T\cosh ^{2}\left[ (\varepsilon -\mu )/2T%
\right] \},  \label{nFd}
\end{equation}%
and the zero-temperature conductivity at energy $\varepsilon $ is
\begin{equation}
\sigma _{zz}(\varepsilon )=\frac{e^{2}\hbar }{2\pi }\sum_{m}v_{z}^{2}(k_{z})%
\left[ 2\text{Im}G_{R}(m,\varepsilon )\right] ^{2}\,,  \label{Sigd1}
\end{equation}%
where $e$ is the electron charge, $v_{z}=\partial \epsilon _{3D}/\partial
k_{z}=2t_{z}d\sin \left( k_{z}d\right) /\hbar $ is the electron velocity,
the sum over the electron quantum numbers $m\equiv \left\{
n,k_{y},k_{z}\right\} $ (excluding spin) is taken in the unit volume, and
the retarded electron Green's function
\begin{equation}
G_{R}=\left[ \varepsilon -\epsilon _{3D}(m)-\Sigma ^{R}(\varepsilon ,m)%
\right] ^{-1}.  \label{GR}
\end{equation}%
Both $\epsilon _{3D}(m)$ and the electron self-energy is independent of $%
k_{y}$, and the summation over $k_{y}$ in Eq. (\ref{Sigd1})\ gives the
factor equal to the LL degeneracy of one conducting layer per spin state $%
N_{LL}=1/2\pi l_{H}^{2}=eB_{z}/2\pi \hbar c$.\cite{CommentNLL} The
interlayer conductivity is now given by a sum over LLs,
\begin{equation}
\sigma _{zz}(\varepsilon )=\sum_{n}\,\sigma _{n}\left( \varepsilon \right) ,
\end{equation}%
where the contribution to $\sigma _{zz}$ from the $n$-th LL is
\begin{equation}
\sigma _{n}=\frac{e^{2}\hbar N_{LL}}{2\pi }\int \frac{dk_{z}}{2\pi }%
v_{z}^{2}(k_{z})\left[ 2\text{Im}G_{R}(n,k_{z},\varepsilon )\right] ^{2}\,.
\label{Sigd2}
\end{equation}

The electron self-energy $\Sigma ^{R}(\varepsilon ,m)=\Sigma
_{n}^{R}(\varepsilon )$ depends only on energy $\varepsilon $ and, possibly,
on LL number $n$, being independent of $k_{z}$ (see Appendix). Substituting
Eq. (\ref{GR}) and performing integration over $k_{z}$, one obtains%
\begin{equation}
\sigma _{n}=\frac{\sigma _{0}\hbar \omega _{c}\Gamma _{0}}{2\pi
t_{z}^{2}\left\vert \text{Im}\Sigma _{n}^{R}(\varepsilon )\right\vert }\text{%
Re}\frac{4t_{z}^{2}-\left( \Delta \varepsilon \right) ^{2}+i\,\Delta
\varepsilon \left\vert \text{Im}\Sigma _{n}^{R}(\varepsilon )\right\vert }{%
\sqrt{4t_{z}^{2}-\left( \Delta \varepsilon -i\,\left\vert \text{Im}\Sigma
_{n}^{R}(\varepsilon )\right\vert \right) ^{2}}},  \label{s0T}
\end{equation}%
where $\Delta \varepsilon \equiv \varepsilon -\epsilon _{2D}\left( n\right)
- $Re$\Sigma _{n}^{R}\left( \varepsilon \right) $, and $\sigma _{0}$ denotes
the interlayer conductivity without magnetic field:%
\begin{equation}
\sigma _{0}=e^{2}\rho _{F}\left\langle v_{z}^{2}\right\rangle \tau
_{0}=2e^{2}N_{LL}t_{z}^{2}d/\hbar ^{2}\omega _{c}\Gamma _{0},  \label{s0}
\end{equation}%
$\rho _{F}=2N_{LL}/\hbar \omega _{c}d$ is the 3D DoS at the Fermi level in
the absence of magnetic field per two spin components, $\tau _{0}=\hbar
/2\Gamma _{0}$ and $\left\langle v_{z}^{2}\right\rangle
=2t_{z}^{2}d^{2}/\hbar ^{2}$.

\medskip

To calculate the electron self-energy $\Sigma _{n}^{R}(\varepsilon )$ we
start from the standard model of 3D strongly anisotropic metals. The
Hamiltonian consists of two terms: $\hat{H}=\hat{H}_{0}+\hat{H}_{I}$. The
first term $\hat{H}_{0}=\sum_{m}\epsilon _{3D}\left( m\right) c_{m}^{+}c_{m}$
describes 3D noninteracting electrons in magnetic field with anisotropic
dispersion given by Eqs. (\ref{ES3D}),(\ref{E2D}). The second term $\hat{H}%
_{I}=\sum_{i}V_{i}\left( r\right) \Psi ^{+}\left( r\right) \Psi \left(
r\right) $ describes the electron interaction with impurities. The
impurities are taken to be point-like with the potential $V_{i}\left(
r\right) =U\delta ^{3}\left( r-r_{i}\right) $ and randomly distributed with
volume concentration $n_{i}$. Without magnetic field the broadening $\Gamma
_{0}=$Im$\Sigma $ of electron levels due to the scattering by impurities in
the Born approximation is $\Gamma _{0}\approx \pi n_{i}U^{2}\rho _{F}$,
where $\rho _{F}$ is the density of electron states (DoS) at the Fermi
level. In Q2D metals in strong magnetic field, $\hbar \omega
_{c}>4t_{z},\Gamma _{0}$, the electron self-energy $\Sigma _{n}^{R}$ depends
on the energy deviation $\Delta \epsilon $ from the $n$-th LL. In the
"non-crossing" approximation,\cite{CommentNC,CommentSigma} schematically
shown in Fig. \ref{FigSE}, the electron Green's function, averaged over
impurity configurations, has the form (see Appendix)%
\begin{equation}
G({\boldsymbol{r}}_{1},{\boldsymbol{r}}_{2},\varepsilon
)=\sum_{n,k_{y},k_{z}}\frac{\Psi _{n,k_{y},k_{z}}^{0\ast }({\boldsymbol{r}}%
_{1})\Psi _{n,k_{y},k_{z}}^{0}({\boldsymbol{r}}_{2})}{\varepsilon -\epsilon
_{3D}\left( n,k_{z}\right) -\Sigma _{n}\left( \varepsilon \right) },
\label{Gg}
\end{equation}%
where $\Psi _{n,k_{y},k_{z}}^{0}({\boldsymbol{r}})$\ are the electron wave
functions in magnetic field without impurities and $\Sigma _{n}\left(
\varepsilon \right) $ is the electron self energy, averaged over impurity
positions and given by the equation
\begin{equation}
\Sigma _{n}(\varepsilon )=n_{i}U/\left[ 1-UG\left( \varepsilon \right) %
\right] .  \label{SER}
\end{equation}%
Here $G\left( \varepsilon \right) \equiv G({\boldsymbol{r}},{\boldsymbol{r}}%
,\varepsilon )$ after the integration over $k_{y}$ and $k_{z}$ is given by%
\begin{equation}
G(\varepsilon )=\sum_{n}\frac{N_{LL}/d}{\sqrt{\left( \varepsilon -\epsilon
_{2D}\left( n\right) -\Sigma _{n}\left( \varepsilon \right) \right)
^{2}-4t_{z}^{2}}}.  \label{Gt}
\end{equation}%
The system of equations (\ref{SER}) and (\ref{Gt}) allows to calculate the
electron self-energy $\Sigma _{n}(\varepsilon )$ numerically.

For weak impurity potential $UG\left( \varepsilon \right) \sim U\rho _{F}\ll
1$ the self-consistent Born approximation (SCBA) is applicable, and Eq. (\ref%
{SER}) reduces to%
\begin{equation}
\Sigma (\varepsilon )\approx n_{i}U+n_{i}U^{2}G\left( \varepsilon \right) .
\label{SEBA}
\end{equation}%
The SCBA is valid when the scattering potential of each impurity is weak
compared to Fermi energy, but the concentration of impurities can be
arbitrary, so that the broadening of electron levels $\Gamma _{0}\approx
n_{i}U^{2}\rho _{F}$ is also arbitrary. In Ref. \cite{WIJETP2011} it was
shown that one needs to apply at least SCBA to obtain qualitatively correct
monotonic growth of interlayer longitudinal MR, which differs in SCBA and in
the non-crossing approximations only by a factor close to unity.

The similar system of SCBA equations for the electron Green's function in
Q2D metals in magnetic field was written previously (see Eqs. (11) and (15)
of Ref. \cite{ShubCondMat1}, Eqs. (11)-(13) in Ref. \cite{Shub}, Eq. (29) in
Ref. \cite{ChampelMineev}, Eq. (17) in Ref. \cite{Gvozd2004} or Eq. (6) in
Ref. \cite{ChMineevComment2006}). There the SCBA equations have not been
solved, and the simple Born approximation was applied to calculate the
interlayer conductivity. In Refs. \cite{ShubCondMat1,Shub} only the limit $%
2\pi t_{z}\gg \hbar \omega _{c}$ has been considered, while in Ref. \cite%
{ChampelMineev} a large electron reservoir was introduced to damp the MQO
and to make the simple Born approximation applicable.\cite{CommentLLSum}

\begin{figure}[tb]
\includegraphics[width=0.49\textwidth]{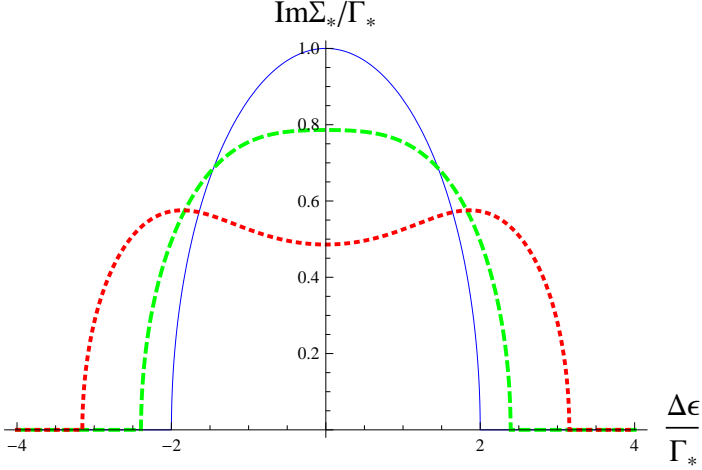}
\caption{{}Imaginary part Im$\Sigma $ of electron self energy as function of
energy deviation $\Delta \protect\epsilon $ from the center of LL at  $%
t_{z}/\Gamma _{\ast }=0$ (blue solid line),$t_{z}/\Gamma _{\ast }=0.5$
(green dashed line), and $t_{z}/\Gamma _{\ast }=1.0$ (red dotted line).}
\label{FigImS}
\end{figure}

\begin{figure}[tb]
\includegraphics[width=0.49\textwidth]{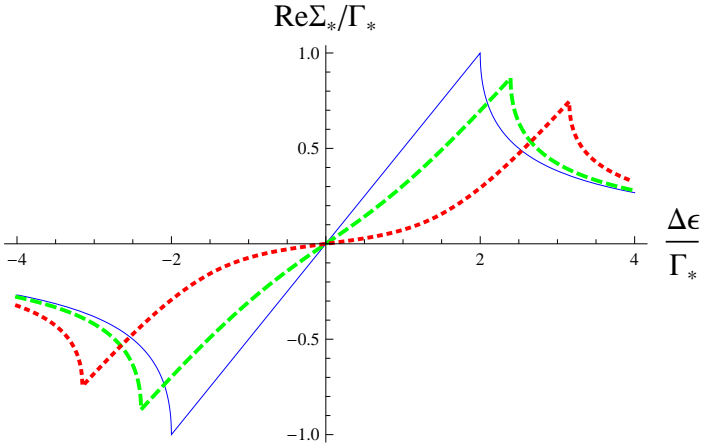}
\caption{{}Real part of electron self energy Re$\Sigma \left( \Delta \protect%
\epsilon \right) $ at $t_{z}/\Gamma _{\ast }=0,0.5,1.0$.}
\label{FigReS}
\end{figure}

Consider the strong magnetic field limit, when $\hbar \omega
_{c}>4t_{z},\Gamma _{0}$ and the LLs do not overlap. Then one can consider
only one LL in the solution of SCBA Eqs. (\ref{Gt}) and (\ref{SEBA}), which
simplify to
\begin{equation}
\Sigma _{\ast }=\frac{n_{i}U^{2}g_{LL}/d}{\sqrt{\left( \Delta \epsilon
-\Sigma _{\ast }\right) ^{2}-4t_{z}^{2}}},  \label{EqS}
\end{equation}%
where $\Sigma _{\ast }\equiv \Sigma _{n}(\varepsilon )-n_{i}U$ and $\Delta
\epsilon \equiv \varepsilon -\epsilon _{2D}\left( n\right) -n_{i}U=\Delta
\varepsilon +$Re$\Sigma _{\ast }^{R}$. At $t_{z}=0$ we obtain the 2D result%
\cite{Ando} for the electron self-energy in SCBA. Note that the scattering
enters Eq. (\ref{EqS}) only in the combination $n_{i}U^{2}g_{LL}/d=\Gamma
_{0}\hbar \omega _{c}$. Therefore, at $t_{z}\rightarrow 0$, $\Gamma _{\ast }=%
\sqrt{\Gamma _{0}\hbar \omega _{c}}$ is the only energy scale in Eq. (\ref%
{EqS}), and naturally Im$\Sigma _{\ast }\sim \sqrt{\Gamma _{0}\hbar \omega
_{c}}\propto \sqrt{B_{z}}$ according to Ref. \cite{Ando}, which leads to the
similar field dependence of background MR:\cite{WIPRB2011,WIJETP2011} $%
R_{zz}\approx R_{zz0}\left\vert \text{Im}\Sigma \left( \mu ,B\right)
\right\vert /\Gamma _{0}\propto \sqrt{B_{z}}$. Eq. (\ref{EqS}) rewrites as
an algebraic equation of the fourth power:%
\begin{equation}
\Sigma _{\ast }^{2}\left[ \left( \Delta \epsilon -\Sigma _{\ast }\right)
^{2}-4t_{z}^{2}\right] =\left( \Gamma _{0}\hbar \omega _{c}\right) ^{2}.
\label{EqS4}
\end{equation}%
Among four solutions of this equation, only one satisfies the physical
requirement $\Sigma _{\ast }\rightarrow 0$ at $\Delta \epsilon \rightarrow
\pm \infty .$ Its solution $\Sigma _{\ast }/\Gamma _{\ast }$ is shown in
Figs. \ref{FigImS},\ref{FigReS} for three different values of $t_{z}/\Gamma
_{\ast }=0,0.5,1.0$. One can see from Fig. 1 that Im$\Sigma _{\ast }\neq 0$
in the finite energy interval of the width $\sim 4\Gamma _{\ast }$ at $%
\Gamma _{\ast }>2t_{z}$ and of the width $\sim 4t_{z}$ at $\Gamma _{\ast
}<2t_{z}$. Re$\Sigma _{n}$ has cusps at the boundaries of this energy
interval.

\begin{figure}[tb]
\includegraphics[width=0.49\textwidth]{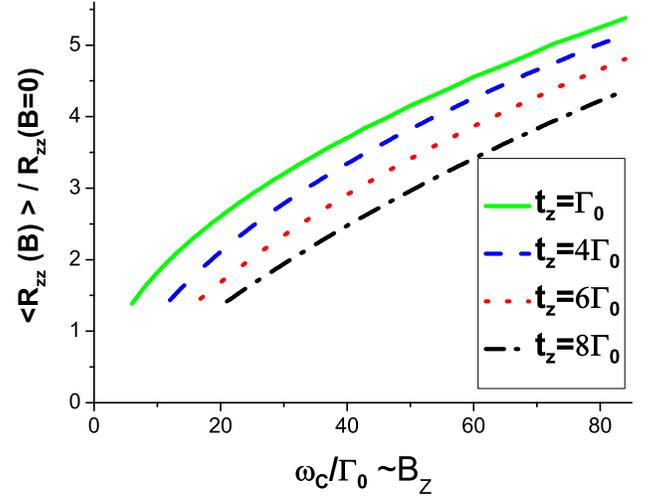}
\caption{Average interlayer magnetoresistance $R_{zz}\left( B_{z}\right) =1/%
\bar{\protect\sigma}_{zz}$ calculated from Eqs. (\protect\ref{s0T}) and (%
\protect\ref{EqS4}) as function of magnetic field at four different values
of $t_{z}/\Gamma _{0}=1$ (solid green line), $t_{z}/\Gamma _{0}=2$ (dashed
blue line), $t_{z}/\Gamma _{0}=3$ (dotted red line) and $t_{z}/\Gamma _{0}=4$
(dash-dotted black line). In the interval $4t_{z}<\hbar \protect\omega %
_{c}\ll \left( 4t_{z}\right) ^{2}/\Gamma _{0}$ the interlayer MR has linear
field dependence.}
\label{FigRzz}
\end{figure}

Now we substitute these solutions into Eq. (\ref{s0T}). The result for
monotonic part of interlayer MR $R_{zz}=\bar{\sigma}_{zz}$ is shown in Fig. %
\ref{FigRzz}, where $\bar{\sigma}_{zz}$ is the conductivity averaged over
period $\hbar \omega _{c}$ of MQO. From this figure we see the crossover
from linear to square-root field dependence of interlayer background MR $%
R_{zz}\left( B_{z}\right) $. The interval of the linear MR is $4t_{z}<\hbar
\omega _{c}\ll \left( 4t_{z}\right) ^{2}/\Gamma _{0}$ and increases with the
increase of $t_{z}/\Gamma _{0}$. The crossover from linear to square-root
dependence of MR is a general feature of quasi-2D metals and already has
been observed in a number of experiments (see, e.g. Refs. \cite%
{Coldea,PesotskiiJETP95}). The square-root dependence $R_{zz}\propto B_{z}$
is obtained at $\hbar \omega _{c}>2\sqrt{\Gamma _{0}\hbar \omega _{c}}%
\gtrsim 4t_{z}$, which is much wider than the applicability region $\hbar
\omega _{c}\gg \Gamma _{0}\gg t_{z}$ of the calculation in Refs. \cite%
{WIPRB2011,WIJETP2011}.
\begin{figure}[tb]
\includegraphics[width=0.49\textwidth]{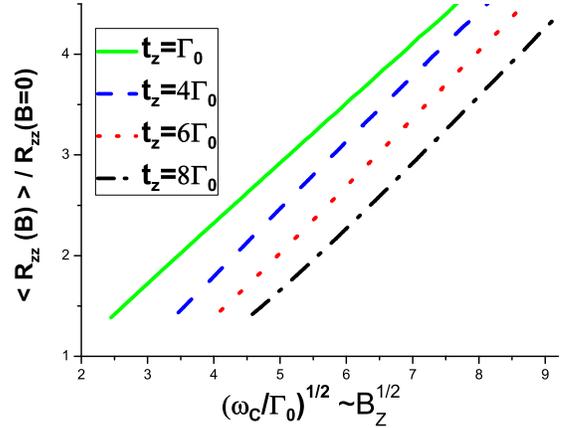}
\caption{Average interlayer magnetoresistance $R_{zz}\left( B_{z}\right) =1/%
\bar{\protect\sigma}_{zz}$ calculated from Eqs. (\protect\ref{s0T}) and (%
\protect\ref{EqS4}) as function of the square root of magnetic field at four
different values of $t_{z}/\Gamma _{0}=1$ (solid green line), $t_{z}/\Gamma
_{0}=2$ (dashed blue line), $t_{z}/\Gamma _{0}=3$ (dotted red line) and $%
t_{z}/\Gamma _{0}=4$ (dash-dotted black line). The linear dependence $%
R_{zz}\left( \protect\sqrt{B_{z}}\right) $ is obtained at $\hbar \protect%
\omega _{c}>2\protect\sqrt{\hbar \protect\omega _{c}\Gamma _{0}}\gtrsim t_{z}
$.  Only in the interval $4t_{z}<\hbar \protect\omega _{c}\ll \left(
4t_{z}\right) ^{2}/\Gamma _{0}$ the interlayer MR has linear field
dependence.}
\label{FigSqrtBz}
\end{figure}

This result on the field dependence of interlayer MR can be
analytically obtained as follows. Eq. (\ref{EqS4}) simplifies to a
quadratic equation at $ \Delta \epsilon =0$, i.e. at the center of
LL, and has two solutions: $\Sigma _{\ast }^{2}=2t_{z}^{2}\pm
\sqrt{4t_{z}^{4}+\left( \Gamma _{0}\hbar \omega _{c}\right)
^{2}}$. The physical solution does not diverge at
$t_{z}\rightarrow \infty $, has nonzero imaginary part and, at
$t_{z}=0$, agrees with the 2D limit described in Refs.
\cite{Ando,Ando1}. All these criteria are satisfied for sign
"$-$", which at $\Delta \epsilon =0$ gives Re$\Sigma _{\ast }=0$
and
\begin{equation}
\left\vert \text{Im}\Sigma _{\ast }\right\vert =\sqrt{\sqrt{%
4t_{z}^{4}+\left( \Gamma _{0}\hbar \omega _{c}\right) ^{2}}-2t_{z}^{2}}.
\label{ImSig1}
\end{equation}%
Substituting this to Eq. (\ref{s0T}) one obtains the expression for the
zero-temperature conductivity in the center of LL, i.e. at $\Delta \epsilon
=0$:%
\begin{equation}
\sigma _{zz}\left( 0\right) \approx 2\sigma _{0}/\pi .  \label{se0}
\end{equation}

\begin{figure}[tb]
\includegraphics[width=0.49\textwidth]{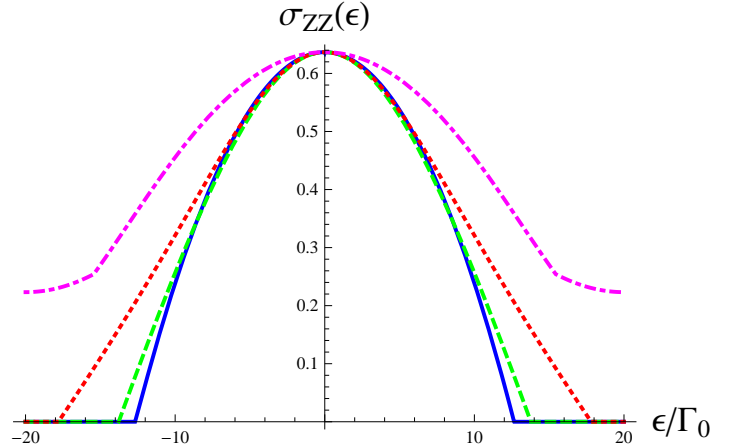}
\caption{Interlayer conductivity $\protect\sigma _{zz}\left(
\epsilon \right) $ calculated from Eqs. (\protect\ref{s0T}) and
(\ref {EqS4}) in the strong-filed limit $\hbar \protect\omega
_{c}/\Gamma _{0}=40$ as function of energy counted from the
nearest LL for four different values of $t_{z}/\Gamma _{0}=0.1$
(solid blue line), $t_{z}/\Gamma _{0}=2$ (dashed green line),
$t_{z}/\Gamma _{0}=5$ (dotted red line), and $t_{z}/\Gamma
_{0}=10$ (dash-dotted magenta line). In the limit $2t_{z}^{2}\gg
\Gamma _{0}\hbar \protect\omega _{c}$ the interlayer conductivity
has approximately a semicircle shape of width $2t_{z}$. In the
opposite limit $\Gamma _{0}\hbar \protect\omega _{c}\gg t_{z}^{2}$
the "conducting band" of each LL has the width $\sim
\protect\sqrt{\Gamma _{0}\hbar \protect\omega _{c}}$.}
\label{FigSEps}
\end{figure}

In the limit $\hbar \omega _{c}>4t_{z}\gg \sqrt{\Gamma _{0}\hbar
\omega _{c}} $ each LL gives an essential contribution to
conductivity in the interval $-2t_{z}\lesssim \Delta \epsilon
\lesssim 2t_{z}$, as follows from Eq. (\ref{EqS4}) and can be seen
from Fig. \ref{FigSEps}, where the conductivity as function of
energy distance from the nearest LL has a dome shape of the width
$\sim 4t_{z}$. The conductivity, averaged over period $\hbar
\omega _{c}$ of MQO, is then given by
\begin{eqnarray}
\bar{\sigma}_{zz} &=&\int_{-\hbar \omega _{c}/2}^{\hbar \omega
_{c}/2}\frac{d\varepsilon }{\hbar \omega _{c}}\sigma _{zz}\left(
\Delta \epsilon \right) \approx \sigma _{zz}\left( 0\right)
\,\frac{\pi }{4}\frac{4t_{z}}{\hbar
\omega _{c}}  \notag  \label{sa} \\
&=&\sigma _{0}\left( 2t_{z}/\hbar \omega _{c}\right) \propto 1/B_{z}.
\label{sav}
\end{eqnarray}%
This predicts a linear background magnetoresistance
$R_{zz}=1/\bar{\sigma}_{zz}\propto B_{z}$ in the interval
$4t_{z}<\hbar \omega _{c}\ll \left( 4t_{z}\right) ^{2}/\Gamma
_{0}$ of magnetic field in quasi-2D strongly anisotropic
compounds. The extra factor $\pi /4$ in Eq. (\ref{sav}) comes from
the integration of a semicircle. Eq. (\ref{sav}) also predicts
stronger dependence of $\bar{\sigma}_{zz}$ on $t_{z}$:
$\bar{\sigma}_{zz}\propto t_{z}^{3}$ unlike the usual dependence
$\bar{\sigma}_{zz}\approx \sigma _{0}\propto t_{z}^{2}$.

In the opposite limit $\hbar \omega _{c}\gg \sqrt{\Gamma _{0}\hbar
\omega _{c}}\gtrsim t_{z}$ the "width of conducting band" from
each LL is $\approx 4 \sqrt{\Gamma _{0}\hbar \omega _{c}}$, and
for background interlayer conductivity one obtains
$\bar{\sigma}_{zz}/\sigma _{0}\approx 2\sqrt{\Gamma _{0}/\hbar
\omega _{c}}\propto 1/\sqrt{B_{z}}$, or
\begin{equation}
\bar{R}_{zz}\left( B\right) /\bar{R}_{zz}\left( 0\right) \approx \sqrt{\hbar
\omega _{c}/4\Gamma _{0}}\propto \sqrt{B_{z}}  \label{RzzSqrt}
\end{equation}
in agreement with Refs. \cite{WIPRB2011,WIJETP2011,WIFNT2011} and
Fig. \ref{FigSqrtBz}.

The magnetic field $B_{cr}$ of the crossover from linear to
square-root dependence of MR $R_{zz}\left( B_{z}\right) $ can be
used to estimate the value of the interlayer transfer integral in
the compound: $t_{z}\sim \sqrt{\Gamma _{0}\hbar \omega _{c}}$,
where $\omega _{c}=eB_{cr}/m^{\ast }c$ corresponds to the
crossover field.

The obtained monotonic dependence of MR originates from MQO but survives at
much higher temperature, when MQO are completely suppressed. According to
Eqs. (\ref{sigmazz})-(\ref{Sigd1}), the temperature smearing of Fermi
distribution function does not influence the monotonic dependence of MR.
Only at much higher temperature, when the electron scattering by phonons
plays the major role in the relaxation of electron momentum, the above
monotonic dependence of MR is weakened. Similarly, the long-range disorder,
not included in the above calculation, produces the local variation of the
Fermi energy along the sample on the scale greater than magnetic length,
which damps the MQO but leaves the monotonic part of MR almost unchanged.
Therefore, even if MQO are not seen in a compound because of strong disorder
or high temperature, the background longitudinal interlayer MR can be
observed and provides useful information about the electronic structure of
this compound. This feature of background MR is similar to that of slow
oscillations in Q2D compounds at $t_{z}>\hbar \omega _{c}$.\cite{SO}

\medskip

To summarize, the calculation of longitudinal interlayer
magnetoresistance in Q2D metals is performed in the strong-field
limit $\hbar \omega _{c}>4t_{z}$. It predicts a linear background
MR $\bar{R}_{zz}\left( B_{z}\right) =1/\bar{\sigma}_{zz}\propto
B_{z}$ in the interval $4t_{z}<\hbar \omega _{c}\ll \left(
4t_{z}\right) ^{2}/\Gamma _{0}$ of magnetic field, where
$\bar{\sigma}_{zz}$ has also unusual dependence on interlayer
transfer integral: $\bar{\sigma}_{zz}\propto t_{z}^{3}$. At
stronger field or at smaller $t_{z}<\Gamma _{0}\hbar \omega _{c}$,
the usual dependence $\bar{\sigma}_{zz}\propto t_{z}^{2}$ is
recovered, and the MR transforms to square-root dependence
$R_{zz}\left( B_{z}\right) \propto \sqrt{B_{z}}$, as was recently
obtained in the limit $t_{z}\ll \Gamma _{0}\ll \hbar \omega _{c}$
in Refs. \cite{WIPRB2011,WIJETP2011,WIFNT2011}. The present
calculation generalizes these results to the region $t_{z}\gtrsim
\Gamma _{0}$. The magnetic field of the crossover from linear to
square-root dependence of $\bar{R}_{zz}\left( B_{z}\right) $
allows to estimate the interlayer transfer integral $t_{z}\sim
\sqrt{\Gamma _{0}\hbar \omega _{c}}$ from experimental data. The
obtained longitudinal MR $\bar{R}_{zz}\left( B_{z}\right) $
explains numerous experiments on interlayer MR in strongly
anisotropic quasi-2D compounds. The measurement of the monotonic
part of longitudinal interlayer MR is much easier than the
measurement of MQO or AMRO, because finite temperature and
long-range crystal imperfections do not affect $\bar{R}_{zz}\left(
B_{z}\right) $ up to much higher temperatures or disorder.
Therefore, the experimental study of longitudinal interlayer
background magnetoresistance is proposed as a simple additional
tool to investigate the electronic structure of strongly
anisotropic quasi-two-dimensional compounds in a wide range of
parameters.

The work was supported by LEA ENS-Landau exchange program and by RFBR.

\section{Appendix}

Now we proof by the method of mathematical induction, that in the
"non-crossing" approximation the electron Green's function, averaged over
impurity configurations, has the form of Eq. (\ref{Gg}):
\begin{equation}
G({\boldsymbol{r}}_{1},{\boldsymbol{r}}_{2},\varepsilon
)=\sum_{n,k_{y},k_{z}}\frac{\Psi _{n,k_{y},k_{z}}^{0\ast }({\boldsymbol{r}}%
_{1})\Psi _{n,k_{y},k_{z}}^{0}({\boldsymbol{r}}_{2})}{\varepsilon -\epsilon
_{3D}\left( n,k_{z}\right) -\Sigma _{n}\left( \varepsilon \right) },
\end{equation}%
with $\Sigma _{n}\left( \varepsilon \right) $ being independent of $k_{y}$
and $k_{z}$.

In the Born approximation with one impurity the electron self energy $\Sigma
_{n}^{B}\left( \varepsilon \right) $ does not depend on $k_{y}$ and $k_{z}$,
but depends on the difference $\Delta \epsilon _{n}\equiv \varepsilon
-\epsilon _{2D}\left( n\right) $:
\begin{equation*}
\Sigma _{n}^{B}\left( \varepsilon \right) \approx U^{2}G_{0}\left(
\varepsilon \right) =\sum_{n}\frac{N_{LL}/d}{\sqrt{\left( \varepsilon
-\epsilon _{2D}\left( n\right) -i\,0\right) ^{2}-4t_{z}^{2}}},
\end{equation*}%
and its real part is an odd function of $\Delta \epsilon _{n}$, Re$\Sigma
_{n}^{B}\left( \Delta \epsilon _{n}\right) =-$Re$\Sigma _{n}^{B}\left(
-\Delta \epsilon _{n}\right) $, while its imaginary part is an even function
of $\Delta \epsilon _{n}$. Hence, in the Born approximation $\Sigma
_{n}\left( \varepsilon ,n,k_{y},k_{z}\right) =\Sigma _{n}\left( \Delta
\epsilon _{n}\right) =\Sigma _{n}\left( \varepsilon \right) $.

Assume that our condition is satisfied when each irreducable self-energy
part contains no more than $j$ impurities. Then the Green's function in
coinciding points depends only on energy, $G_{j}\left( \varepsilon
,r,r\right) =G_{j}\left( \varepsilon \right) $:
\begin{eqnarray}
G_{j}\left( \varepsilon ,r,r\right) &=&\sum_{n,k_{y},k_{z}}\frac{\Psi
_{n,k_{y},k_{z}}^{0\ast }(r)\Psi _{n,k_{y},k_{z}}^{0}(r)}{\varepsilon
-\epsilon _{3D}\left( n,k_{z}\right) -\Sigma _{n,j}\left( \varepsilon
\right) }  \notag \\
&=&\sum_{n,k_{z}}\frac{N_{LL}/d}{\varepsilon -\epsilon _{3D}\left(
n,k_{z}\right) -\Sigma _{n,j}\left( \varepsilon \right) }.  \label{Gj}
\end{eqnarray}%
When we include one more impurity to the irreducable electron self-energy
and average over its position, it remains independent of $k_{y},k_{z}$ and
depends only on $\Delta \epsilon _{n}$:
\begin{equation*}
\Sigma _{n,j+1}=Un_{i}/\left[ 1-UG_{j}\left( \Delta \epsilon _{n}\right) %
\right] .
\end{equation*}%
The Green's function $G_{j+1}\left( \varepsilon ,r_{1},r_{2}\right) $ also
keeps the form of Eq. (\ref{Gg}). To show this, let us find the function
\begin{eqnarray*}
&&G_{j+1}\left( \varepsilon ,r_{1},r_{2}\right) =G_{0}\left( \varepsilon
,r_{1},r_{2}\right) \\
&&+\int d^{3}r\frac{UG_{0}\left( r_{1},r\right) G_{0}\left( r,r_{2}\right) }{%
1-UG_{j}\left( \varepsilon ,r,r\right) } \\
&&+\frac{\int d^{3}r\int d^{3}r^{\prime }UG_{0}\left( r_{1},r\right)
G_{0}\left( r,r^{\prime }\right) G_{0}\left( r^{\prime },r_{2}\right) }{%
\left[ 1-UG_{j}\left( \varepsilon ,r,r\right) \right] \left[ 1-UG_{j}\left(
\varepsilon ,r^{\prime },r^{\prime }\right) \right] }+.. \\
&=&G_{0}\left( \varepsilon ,r_{1},r_{2}\right) +\frac{U}{1-UG_{j}\left(
\varepsilon \right) }\times \\
&&\times \sum_{n,k_{y},k_{z}}\frac{\Psi _{n,k_{y},k_{z}}^{0}(r_{1})}{%
\varepsilon -\epsilon _{3D}\left( n,k_{z}\right) -\Sigma _{0,n}\left(
\varepsilon \right) } \\
&&\times \sum_{n^{\prime },k_{y}^{\prime },k_{z}^{\prime }}\frac{\Psi
_{n^{\prime },k_{y}^{\prime },k_{z}^{\prime }}^{0\ast }({\boldsymbol{r}}_{2})%
}{\varepsilon -\epsilon _{3D}\left( n^{\prime },k_{z}^{\prime }\right)
-\Sigma _{0,n^{\prime }}\left( \varepsilon \right) } \\
&&\times \int d^{3}r\Psi _{n,k_{y},k_{z}}^{0\ast }(r)\Psi _{n^{\prime
},k_{y}^{\prime },k_{z}^{\prime }}^{0}({\boldsymbol{r}})+..
\end{eqnarray*}%
The integration over $d^{3}r$ in the last line gives $\delta \left(
n-n^{\prime }\right) \delta \left( k_{y}-k_{y}^{\prime }\right) \delta
\left( k_{z}-k_{z}^{\prime }\right) $, and the function $G_{j+1}\left(
\varepsilon ,r_{1},r_{2}\right) $ becomes
\begin{eqnarray}
&&G_{0}\left( \varepsilon ,r_{1},r_{2}\right) +\frac{Un_{i}}{1-UG_{j}\left(
\varepsilon \right) }\times  \notag \\
&&\sum_{n,k_{y},k_{z}}\frac{\Psi _{n,k_{y},k_{z}}^{0}(r_{1})\Psi
_{n,k_{y},k_{z}}^{0\ast }({\boldsymbol{r}}_{2})}{\left[ \varepsilon
-\epsilon _{3D}\left( n,k_{z}\right) -\Sigma _{0,n}\left( \varepsilon
\right) \right] ^{2}}+..  \notag \\
&=&\sum_{n,k_{y},k_{z}}\frac{\Psi _{n,k_{y},k_{z}}^{0}(r_{1})\Psi
_{n,k_{y},k_{z}}^{0\ast }({\boldsymbol{r}}_{2})}{\varepsilon -\epsilon
_{3D}\left( n,k_{z}\right) -\Sigma _{0,n}\left( \varepsilon \right) }  \notag
\\
&&\times \sum_{i=0}^{\infty }\left( \frac{Un_{i}/\left[ 1-UG_{j}\left(
\varepsilon \right) \right] }{\varepsilon -\epsilon _{3D}\left(
n,k_{z}\right) -\Sigma _{0,n}\left( \varepsilon \right) }\right) ^{i}  \notag
\\
&=&\sum_{n,k_{y},k_{z}}\frac{\Psi _{n,k_{y},k_{z}}^{0}(r_{1})\Psi
_{n,k_{y},k_{z}}^{0\ast }({\boldsymbol{r}}_{2})}{\varepsilon -\epsilon
_{3D}\left( n,k_{z}\right) -\Sigma _{j+1,n}\left( \varepsilon \right) },
\label{Gj1}
\end{eqnarray}%
where the new self-energy part
\begin{equation*}
\Sigma _{j+1,n}\left( \varepsilon \right) =\Sigma _{0,n}\left( \varepsilon
\right) +Un_{i}/\left[ 1-UG_{j}\left( \varepsilon \right) \right]
\end{equation*}%
depends only on energy $\varepsilon $ and on LL number $n$. The Green's
function in Eq. (\ref{Gj1}) is again of the form of Eq. (\ref{Gg}).

The difference of impurity scattering to the same and to other LLs
may give additional dependence of $\Sigma _{n}\left( \varepsilon
\right) $ on LL number $n$, which requires further study. If
$\Sigma _{n,0}\left( \varepsilon \right) $ depends only on $\Delta
\epsilon _{n}$, then $G_{j}\left( \varepsilon \right) $ entering
above formulas also depends only on $\Delta \epsilon _{n}$, and
$\Sigma _{n,j+1}\left( \varepsilon \right) $ is also a function of
$\Delta \epsilon _{n}$ only. If $\Sigma _{n,0}\left( \varepsilon
\right) $ depends only on $\varepsilon $ but not on $n$, and if
the impurity scattering to the same and to other LLs has the same
matrix element, then $G_{j}\left( \varepsilon \right) $ also
depends only on $\varepsilon $, and so does $\Sigma _{n,j+1}\left(
\varepsilon \right) $.
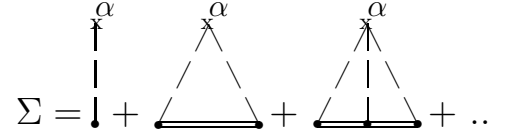
\begin{figure}[tb]
{\small 
}
\par
\begin{center}
{\small
\begin{picture}(120,55)
\put(0, 5){\Large{$\Sigma =$}}
\multiput(30,5)(0,14){3}{\line(0,1){10}} \put(30,
46){\large{$\alpha $}} \put(28, 41){x} \put(30, 5){\circle*{3}}
\put(36, 5){\Large{+}} \multiput(54,5)(7,14){3}{\line(1,2){5}}
\multiput(92,5)(-7,14){3}{\line(-1,2){5}} \put(73,
46){\large{$\alpha $}} \put(70, 41){\small{x}} \put(54,
5){\circle*{3}} \put(92, 5){\circle*{3}} \put(54,
6){\line(1,0){38}} \put(54, 4){\line(1,0){38}} \put(96,
5){\Large{+}} \multiput(114,5)(7,14){3}{\line(1,2){5}}
\multiput(133,5)(0,14){3}{\line(0,1){10}}
\multiput(152,5)(-7,14){3}{\line(-1,2){5}} \put(133,
46){\large{$\alpha $}} \put(130, 41){\small{x}} \put(114,
5){\circle*{3}} \put(152, 5){\circle*{3}} \put(133,
5){\circle*{3}} \put(114, 6){\line(1,0){38}} \put(114,
4){\line(1,0){38}} \put(156, 5){\Large{+ ..}}
\end{picture}
}
\par
{\small 
}
\end{center}
\caption{ The set of diagrams for the irreducible self-energy, corresponding
to the self-consistent one-site approximation. The double solid line
symbolizes the exact electron Green's function. }
\label{FigSE}
\end{figure}


\begin{thebibliography}{99}
\bibitem{MQORev} J. Wosnitza, \textit{Fermi Surfaces of Low-Dimensional
Organic Metals and Superconductors} (Springer-Verlag, Berlin, 1996); J.
Singleton, Rep. Prog. Phys. \textbf{63}, 1111 (2000).

\bibitem{OMRev} T.~Ishiguro, K.~Yamaji and G.~Saito, \emph{Organic
Superconductors}, 2nd Edition, Springer-Verlag, Berlin, 1998.

\bibitem{MarkReview2004} M.V. Kartsovnik, Chem. Rev. \textbf{104}, 5737
(2004).

\bibitem{KartPeschReview} M. V. Kartsovn\u{\i}k and V. G. Peschansky, Low
Temp. Phys. \textbf{31}, 185 (2005) [Fiz. Nizk. Temp. \textbf{31}, 249
(2005)].

\bibitem{HusseyNature2003} N. E. Hussey, M. Abdel-Jawad, A. Carrington, A.
P. Mackenzie and L. Balicas, Nature \textbf{425}, 814 (2003).

\bibitem{AbdelNature2006} M. Abdel-Jawad, M. P. Kennett, L. Balicas, A.
Carrington, A. P. Mackenzie, R. H. McKenzie \& N. E. Hussey, Nature Phys.
\textbf{2}, 821 (2006).

\bibitem{ProustNature2007} Nicolas Doiron-Leyraud, Cyril Proust, David
LeBoeuf, Julien Levallois, Jean-Baptiste Bonnemaison, Ruixing Liang, D. A.
Bonn, W. N. Hardy, Louis Taillefer, Nature \textbf{447}, 565 (2007).

\bibitem{AbdelPRL2007AMRO} M. Abdel-Jawad, J. G. Analytis, L. Balicas et
al., Phys. Rev. Lett. \textbf{99}, 107002 (2007).

\bibitem{McKenzie2007} Malcolm P. Kennett and Ross H. McKenzie, Phys. Rev. B
\textbf{76}, 054515 (2007).

\bibitem{DVignolle2008} B. Vignolle, A. Carrington, R. A. Cooper, M. M. J.
French, A. P. Mackenzie, C. Jaudet, D. Vignolles, Cyril Proust \& N. E.
Hussey, Nature \textbf{455}, 952 (2008).

\bibitem{HelmNd2009} T. Helm,1 M.V. Kartsovnik, M. Bartkowiak, N. Bittner,
M. Lambacher, A. Erb, J. Wosnitza, and R. Gross, Phys. Rev. Lett. \textbf{103%
}, 157002 (2009).

\bibitem{HelmNd2010} T. Helm, M.V. Kartsovnik, I. Sheikin et al., Phys. Rev.
Lett. \textbf{105}, 247002 (2010).

\bibitem{BaFeAs2011} Taichi Terashima, Nobuyuki Kurita, Megumi Tomita et
al., Phys. Rev. Lett. \textbf{107}, 176402 (2011).

\bibitem{Graf2012} D. Graf, R. Stillwell, T. P. Murphy et al., Phys. Rev. B.
\textbf{85}, 134503 (2012).

\bibitem{Kuraguchi2003} M. Kuraguchi et al., Synth. Met. \textbf{133-134},
113 (2003).

\bibitem{AMB6} Hiroyuki Takeya and Mohammed ElMassalami, Phys. Rev. B
\textbf{84}, 064408 (2011).

\bibitem{Abrik} A.A. Abrikosov, \textit{Fundamentals of the theory of metals},
 North-Holland, 1988.

\bibitem{Shoenberg} Shoenberg D. \textquotedblright Magnetic oscillations in
metals\textquotedblright , Cambridge University Press 1984.

\bibitem{Ziman} J. M. Ziman, \textit{Principles of the Theory of Solids},
Cambridge Univ. Press 1972.

\bibitem{KartsAMRO1988} M.V. Kartsovnik, P. A. Kononovich , V. N. Laukhin
and I. F. Shchegolev, JETP Lett. \textbf{48}, 541 (1988).

\bibitem{Yam} K. Yamaji, J. Phys. Soc. Jpn. \textbf{58}, 1520 (1989).

\bibitem{Yagi1990} R. Yagi, Y. Iye, T. Osada, S. Kagoshima, J. Phys. Soc.
Jpn. \textbf{59}, 3069 (1990).

\bibitem{SO} M.V. Kartsovnik, P.D. Grigoriev, W. Biberacher, N.D. Kushch, P.
Wyder, Phys. Rev. Lett. \textbf{89}, 126802 (2002).

\bibitem{Shub} P.D. Grigoriev, Phys. Rev. B \textbf{67}, 144401 (2003)
[arXiv:cond-mat/0204270].

\bibitem{PhSh} P.D. Grigoriev, M.V. Kartsovnik, W. Biberacher, N.D. Kushch,
P. Wyder, Phys. Rev. B \textbf{65}, 60403(R) (2002).

\bibitem{CommentMaltsev} Some nonzero longitudinal magnetoresistance may
appear in the $\tau $-approximation only in some special cases of
complicated FS and particular directions of magnetic field, when there are
singular electron trajectories.\cite{Maltsev1997}

\bibitem{Coldea} A.I. Coldea et al., Phys. Rev. B \textbf{69}, 085112 (2004).

\bibitem{PesotskiiJETP95} R.B. Lyubovskii, S.I. Pesotskii, A. Gilevskii and
R.N. Lyubovskaya, JETP \textbf{80}, 946 (1995) [Zh. Eksp. Teor. Fiz. \textbf{%
107}, 1698 (1995)].

\bibitem{Zuo1999} F. Zuo, X. Su, P. Zhang, J. S. Brooks, J. Wosnitza, J. A.
Schlueter, Jack M. Williams, P. G. Nixon, R. W. Winter, and G. L. Gard,
Phys. Rev. B \textbf{60}, 6296 (1999).

\bibitem{W3} J. Hagel, J. Wosnitza, C. Pfleiderer, J. A. Schlueter, J.
Mohtasham, and G. L. Gard, Phys. Rev. B \textbf{68}, 104504 (2003).

\bibitem{W4} J.Wosnitza, Journal of Low Temperature Physics 146, 641 (2007).

\bibitem{Incoh2009} M. V. Kartsovnik, P. D. Grigoriev, W. Biberacher, and N.
D. Kushch, Phys. Rev. B \textbf{79}, 165120 (2009).

\bibitem{Kang} W. Kang, Y. J. Jo, D. Y. Noh, K. I. Son, and Ok-Hee Chung,
Phys. Rev. B \textbf{80}, 155102 (2009).

\bibitem{Wosnitza2002} J.Wosnitza, J. Hagel, J. S. Qualls, J. S. Brooks, E.
Balthes, D. Schweitzer, J. A. Schlueter, U. Geiser, J. Mohtasham, R. W.
Winter, et al., Phys. Rev. B 65, 180506(R) (2002).

\bibitem{Gvozd2007} V. M. Gvozdikov,Phys. Rev. B \textbf{76}, 235125 (2007).

\bibitem{Lundin2003} Urban Lundin and Ross H. McKenzie, Phys. Rev. B \textbf{%
68}, 081101(R) (2003).

\bibitem{Ho} A. F. Ho and A. J. Schofield, Phys. Rev. B 71, 045101 (2005).

\bibitem{WIPRB2011} P.D. Grigoriev, Phys. Rev. B \textbf{83}, 245129 (2011).

\bibitem{WIJETP2011} P.D. Grigoriev, JETP Lett. \textbf{94}, 47 (2011).

\bibitem{WIFNT2011} P.D. Grigoriev, Fiz. Nizk. Temp. \textbf{37}, 930 (2011)
[Low Temp. Phys. \textbf{37}, 738 (2011)].

\bibitem{MosesMcKenzie1999} P. Moses and R.H. McKenzie, Phys. Rev. B \textbf{%
60}, 7998 (1999).

\bibitem{Mahan} G. Mahan \textquotedblright Many-Particle
Physics\textquotedblright , 2nd ed., Plenum Press, New York, 1990.

\bibitem{CommentChemPotOsc} The MQO of the chemical potential $\mu $ can be
ignored because they are usually strongly suppressed by magnetostriction of
the sample, as was first shown in Ref. \cite{Magnetostriction}.

\bibitem{CommentNLL} This notation $N_{LL}$ denotes the area electron
concentration on one LL. It coincides with $N_{LL}$ in Ref. \cite{WIPRB2011}
and differs from the same notation in Ref. \cite{Shub} by the factor $d$,
because in Ref. \cite{Shub} $N_{LL}$ denoted the volume LL degeneracy.

\bibitem{CommentNC} In the 3D case the diagrams with the intersections of
impurity lines are small by the parameter $n_{i}/n_{e}\ll 1$, where $n_{i}$
and $n_{e}$ are the volume impurity and electron concentrations. In 2D case
in magnetic field these crossing diagrams are not small by this parameter.
However, the calculations of the DoS in Refs. \cite{Ando1,Brezin} show that
these diagrams give a small correction to the result of "non-crossing"
approximation, they describe the exponentially small tails of the DoS. In
quasi-2D case, which is intermediate between 3D and 2D case, the diagrams
with intersections of impurity lines can also be neglected.

\bibitem{CommentSigma} The spatial inhomogeniety of the electron density in
$z$-direction due to the layered crystal structure also leads to some
corrections in the electron self energy, which are equivalent to additional
integration over the strength of the impurity potential.\cite{Imp,Burmi}

\bibitem{ShubCondMat1} P.D. Grigoriev, M.V. Kartsovnik, W. Biberacher, P.
Wyder, arXiv:cond-mat/0108352 (unpublished); P.D. Grigoriev, Ph.D. thesis
Univ. Konstanz (2002).

\bibitem{ChampelMineev} T. Champel and V. P. Mineev, Phys. Rev. B
\textbf{66},195111 (2002).

\bibitem{Gvozd2004} V. M. Gvozdikov, Phys. Rev. B \textbf{70}, 085113 (2004).

\bibitem{ChMineevComment2006} T. Champel and V. P. Mineev, Phys. Rev. B
\textbf{74}, 247101 (2006).

\bibitem{CommentLLSum} The summation over LLs, performed in Refs.
\cite{Shub,ShubCondMat1,ChampelMineev,Gvozd2004,ChMineevComment2006} using the
Poisson summation formula to derive the SCBA equations, assumes that $\Sigma
_{n}\left( \varepsilon \right) $ is independent on LL number $n$, which
requires additional proof in Q2D limit.

\bibitem{Ando} Tsunea Ando, J. Phys. Soc. Jpn. \textbf{36}, 1521 (1974).

\bibitem{Maltsev1997} A.Y. Mal'tsev, JETP \textbf{85}, 934 (1997) [Zh. Exp.
Teor. Fiz. \textbf{112}, 1710 (1997)].

\bibitem{Magnetostriction} N.E.Alekseevskii and V.I.Nizhanovskii, Zh. Eksp.
Teor. Fiz. \textbf{61}, 1051 (1985) [JETP \textbf{88}, 1771 (1985)].

\bibitem{Ando1} Tsunea Ando, J. Phys. Soc. Jpn. \textbf{37}, 622 (1974).

\bibitem{Brezin} E. Brezin, D.I. Gross, C. Itzykson. Nucl. Phys. B
\textbf{235}, 24 (1984).

\bibitem{Imp} A.M. Dyugaev, P.D. Grigor'ev, Yu.N. Ovchinnikov, JETP Letters
\textbf{78}, 148 (2003).

\bibitem{Burmi} I.S. Burmistrov, M.A. Skvortsov, JETP Lett. \textbf{78}, 156
(2003).
\end{thebibliography}
\end{document}